\documentclass[12pt]{article}
\usepackage{amssymb}
\usepackage{epsfig}
\usepackage{float}
\usepackage{graphicx}
\usepackage{amsmath}
\newcommand{\nua}[1]{\ensuremath{\rlap
           {\kern-2.5pt\ensuremath
           {\overset{\scriptscriptstyle(-)}{\phantom{\nu}}}}
           {\ensuremath{{\nu}_{#1}}}}}
\begin{document}

\begin{center}
{\bf Neutrinos: Majorana or Dirac?}
\end{center}
\begin{center}
S. M. Bilenky
\end{center}
\begin{center}
{\em  Joint Institute for Nuclear Research, Dubna, R-141980,
Russia\\}
\end{center}
\begin{abstract}
Are neutrinos with definite masses  Majorana or Dirac particles? This is one of the most fundamental problem of the modern neutrino physics. The solution of this problem could be crucial for understanding of the origin of small neutrino masses. We will review here basic arguments in favor of the Majorana nature of massive neutrinos. The phenomenological theory of $0\nu\beta\beta$-decay is briefly discussed and  recent experimental data and sensitivity of future experiments are presented.
\end{abstract}

\section{Introduction}

The origin of small neutrino masses, discovered in neutrino oscillation experiments, is the major problem of the modern neutrino physics. From neutrino oscillation data five neutrino oscillation parameters (two neutrino mass-squared differences and three mixing angles) were inferred with accuracies (5-10) \%. The aim of current and future experiments is to improve accuracy of the measurement of these parameters and to answer the following basic questions
\begin{itemize}
  \item What is the character of the neutrino mass spectrum (Normal or Inverted Ordering?)
  \item What is the value of the $CP$ phase $\delta$?
  \item How many neutrinos with definite masses $\nu_{i}$ exist in nature? Is the number of $\nu_{i}$ is equal to the number of flavor neutrinos $\nu_{l}$ ($l=e,\mu,\tau$)  or larger  (in this case exist additional sterile neutrinos)?
\item What is the nature of neutrinos with definite masses? Are they  Majorana or Dirac particles?
\end{itemize}
The solution of these problems will be extremely important for the understanding of the origin of neutrino masses. We will discuss here the problem of the neutrino nature, which, apparently, is the most fundamental one.

  Neutrinos with definite masses are Dirac particles if the total lepton number $L$ is conserved. In this case neutrino $\nu_{i}$ and antineutrino $\bar\nu_{i}$ have the same mass  ($CPT$) and different lepton numbers ($L(\nu_{i})=-L(\bar\nu_{i})=1$). Neutrinos with definite masses are Majorana particles if there no conserved lepton number (i.e there is no conserved quantum number which allows to distinguish neutrino or antineutrino).

{\em There is a general  believe that   neutrinos   are Majorana particles}.  We will start with  some  general argument in favor of Majorana neutrinos. The famous two-component theory of a massless neutrino was proposed in 1957 by Landau \cite{Landau:1957tp}, Lee and Yang \cite{Lee:1957qr} and Salam \cite{Salam:1957st} and  was confirmed in
in the classical   Goldhaber at al. experiment on the measurement of the neutrino helicity \cite{Goldhaber:1958nb}. The two-component Weil field $\nu_{L}(x)$ is the  simplest possibility for massless neutrino: two degrees of freedom (instead of four in the case of four-component Dirac neutrino).\footnote{Notice that for massless two-component neutrino and $V-A$ interaction there is no difference between Dirac and Majorana cases.}

We know now that neutrinos have small masses. For neutrino with a mass {\em Majorana field is a simplest, most economical possibility}: two degrees of freedom (left-handed and right-handed neutrinos). The Standard Model teaches us that the nature chooses the simplest possibilities. It looks very plausible that also in the case of neutrino with mass a simplest Majorana possibility is realized.

Neutrino masses are many orders of magnitude smaller then masses of leptons and quarks. It is very
unlikely that neutrino masses are of the same Standard Model (SM) Higgs origin as masses of other fundamental fermions. If we assume that Standard Model  neutrinos are two-component massless particles in this case neutrino masses are generated by a new, beyond the SM mechanism. The method of the effective Lagrangian allows us to describe effects of a beyond the SM physics in the electroweak region. There exist only one $SU_{L}(2)\times U_{Y}(1)$ invariant effective Lagrangian which generates a neutrino mass term \cite{Weinberg:1979sa}. We can built such a Lagrangian only if we assume that the total lepton number $L$ is not conserved. After spontaneous symmetry breaking this effective Lagrangian generates Majorana neutrinos with definite masses, three-neutrino mixing  and seesaw-type suppression of neutrino masses with respect to Standard Model masses of lepton and quarks. This is a simplest and most plausible, beyond the SM possibility to generate neutrino masses and mixing.\footnote{``Simplicity is a guide to the theory choice" A. Einstein.}

 There are many models
which (after heavy fields are  integrated out) lead to the Weinberg effective Lagrangian (or its generalization) and the Majorana mass term (see review \cite{Cai:2017jrq}). In spite all such models were proposed to explain smallness of neutrino masses, values of masses  can not be predicted (many unknown parameters are involved). There are, however, two  general features which are common to all models,  based on the assumption that SM neutrinos are massless particles and that beyond SM particles, responsible for generation of neutrino masses, are heavy:
\begin{enumerate}
  \item Neutrinos with definite masses are Majorana particles.
  \item The number of neutrinos with definite masses is equal to the number of flavor neutrinos (there is no sterile neutrinos).
 \end{enumerate}
The problem of sterile neutrinos, which started about 25 years with first data of the LSND experiment \cite{Aguilar:2001ty},  is still open. We will briefly discuss it later. The study of neutrinoless double $\beta$-decay of some even-even nuclei ($0\nu\beta\beta$-decay) is the most sensitive way which could allow us to discover the non conservation of the total lepton number $L$ and to reveal the Majorana nature of neutrinos with definite masses. We will briefly discuss the phenomenological theory of this process and present the latest data.

In conclusion we would like to stress that in spite of  strong arguments in favor of Majorana neutrinos  the possibility of Dirac neutrinos (of the Standard Model or beyond the SM origin) is not excluded.  The observation  of the $0\nu\beta\beta$-decay will allow to exclude this, apparently, artificial possibility.

\section{On the Higgs Mechanism of the Generation of  Fermion Masses}
From the discovery of neutrino oscillations,  measurement of neutrino masses in tritium experiments and cosmological data  follows that neutrino masses are many orders of magnitudes smaller than masses of leptons and quarks. From this basic experimental fact we can conclude that  it is very plausible that {\em neutrino masses and masses of other fundamental fermions are of different origin}. We will discuss a possible origin of neutrino masses later. In this introductory section we will consider the standard Higgs mechanism of the generation of fermion masses.

The standard Higgs  mechanism of the generation of
 fermion masses  is based on the assumption that in the total Lagrangian there are $SU_{L}(2)\times U_{Y}(1)$ invariant Yukawa interactions. For the charged leptons the Yukawa interaction has the form
\begin{equation}\label{Yukawa}
    \mathcal{L}_{Y}(x)=-\sqrt{2}\sum_{l_{1},l_{2}}\bar\psi_{l_{1}L}(x)
    Y_{l_{1}l_{2}}l'_{2R}(x)\phi(x)+\mathrm{h.c.}.
\end{equation}
Here
\begin{eqnarray}\label{1Yukawa}
\psi_{lL}(x)=\left(
\begin{array}{c}
\nu'_{lL}(x) \\
l'_{L}(x) \\
\end{array}
\right)~~ (l=e,\mu,\tau),\quad\phi(x)=\left(
\begin{array}{c}
\phi_{+}(x)
 \\
\phi_{0}(x) \\
\end{array}
\right)
\label{ldoub}
\end{eqnarray}
are lepton and Higgs doublets,  $l'_{R}(x)$ is a right-handed lepton singlet field and $Y$ is a $3\times3$ dimensionless complex matrix. The requirements of the $SU_{L}(2)\times U_{Y}(1)$ invariance do not put any  constraints on the matrix $Y$.

Charged lepton masses are generated after the spontaneous symmetry breaking. Let us introduce the hermitian field of the neutral Higgs particles $H(x)$   and choose the Higgs doublet in the form (the unitary gauge)
\begin{eqnarray}\label{spontsb}
\phi(x)= \left(
\begin{array}{c}
0 \\
\frac{v+H(x)}{\sqrt{2}}\\
\end{array}
\right).
\end{eqnarray}
Here $v=(\sqrt{2}~G_{F})^{-1/2}\simeq 246~\mathrm{GeV}$ ($G_{F}$ is the Fermi constant) is the Higgs vacuum expectation value (vev). With such a choice the local $SU_{L}(2)\times U_{Y}(1)$ gauge symmetry will be spontaneously broken.

From (\ref{Yukawa}) and (\ref{spontsb}) for the Yukawa Lagrangian we find the following expression
\begin{equation}\label{Yukawa1}
\mathcal{L}_{Y}(x)=-\sum_{l_{1},l_{2}}\bar l'_{1L}(x)
    Y_{l_{1}l_{2}}l'_{2R}(x)(v+H(x))+\mathrm{h.c.}
\end{equation}
The $3\times3$ complex matrix $Y$ can be diagonalized by the biunitary transformation:
\begin{equation}\label{Yukawa3}
Y=V_{L}~y~V^{\dag}_{R},
\end{equation}
where $V_{L}$ and $V_{R}$ are unitary matrices and
\begin{equation}\label{Yukawa4}
y_{l'l}=y_{l}~\delta_{l'l},~~~y_{l}\geq 0\quad (l,l'=e,\mu,\tau).    \end{equation}
From (\ref{Yukawa1}) and (\ref{Yukawa3}) we find that the Yukawa Lagrangian takes the form
\begin{equation}\label{Yukawa5}
 \mathcal{L}_{Y}(x)=-\sum_{l=e,\mu,\tau}m_{l}~\bar l(x) l(x) -\sum_{l=e,\mu,\tau}y_{l}~\bar l(x) l(x)~H(x)
\end{equation}
Here $l(x)= l_{L}(x)+ l_{R}(x)$  is the field of the leptons $l^{\pm}$ ($l=e,\mu,\tau$) with the mass
\begin{equation}\label{Yukawa6}
m_{l}=y_{l}~v.
\end{equation}
The fields $l_{L,R}(x)$ are connected with primed fields $l'_{L,R}(x)$ by the unitary transformations
\begin{equation}\label{Yukawa7}
 l_{L}(x)=\sum_{l_{1}} (V^{\dag}_{L})_{ll_{1}}l'_{1L}(x),\quad
l_{R}(x)=\sum_{l_{1}} (V^{\dag}_{R})_{ll_{1}}l'_{1R}(x)
\end{equation}
Thus, Yukawa interaction (\ref{Yukawa}) after spontaneous symmetry breaking generates the standard mass term of the charged leptons and the Lagrangian of the interaction of
charged leptons and the Higgs field. The Yukawa constants of this interaction are determined by the lepton mass and are given by
\begin{equation}\label{Yukawa8}
  y_{l}=\frac{m_{l}}{v}.
\end{equation}
The SM masses of up and down quarks are generated in the same way as charged lepton's masses. For masses of quarks we find
\begin{equation}\label{Yukawa9}
m_{q}=y_{q}~v~~~~(q=u,d,c,s,t,b).
\end{equation}
The Yukawa constants of the quark-Higgs interaction are determined by quark masses and are given by
\begin{equation}\label{Yukawa10}
y_{q}=\frac{m_{q}}{v}.
\end{equation}
Let us notice that Yukawa constants for different channels were determined from the LHC study of the decay of the Higgs boson into fermion-antifermion pairs (see \cite{Tanabashi:2018oca}). The obtained data are in a good agreement with the SM predicted values (\ref{Yukawa8}) and (\ref{Yukawa10}). From our point of view this is an important argument in favor of the Higgs mechanism of quark and lepton masses generation.

Formally neutrino masses can also be generated by the standard Higgs mechanism. In fact, let us assume that  into the total Lagrangian
enters the $SU_{L}(2)\times U_{Y}(1)$ invariant Yukawa interaction
\begin{equation}\label{YukawaNu}
\mathcal{L}^{\nu}_{Y}(x)    =-\sqrt{2}\sum_{l_{1},l_{2}}\bar\psi_{l_{1}L}(x)
    Y^{\nu}_{l_{1}l_{2}}\nu'_{l_{2}R}(x)\tilde{\phi}(x)+\mathrm{h.c.}
\end{equation}
where $\tilde{\phi}=i\tau_{2}~\phi^{*}$ is the conjugated Higgs doublet and right-handed fields $\nu'_{lR}$ are singlets. Thus,  we need to assume that not only left-handed flavor neutrino fields $\nu'_{lL}$ but also  right-handed fields $\nu'_{lR}$ are SM fields.

After the spontaneous symmetry breaking we find
\begin{equation}\label{YukawaNu2}
\mathcal{L}^{\nu}_{Y}(x)=-\sum_{l_{1},l_{2}}\bar\nu'_{l_{1}L}(x)
    Y^{\nu}_{l_{1}l_{2}}\nu'_{l_{2}R}(x)(v+H(x))+\mathrm{h.c.}
\end{equation}
The proportional to $v$ term of this expression is the  neutrino mass term. After the standard diagonalization of  $3\times 3$ matrix $Y^{\nu}$ for the neutrino mass term we obtain the following expression
\begin{equation}\label{YukawaNu5}
\mathcal{L}^{\nu}(x) =-\sum^{3}_{i=1}m_{i}~(\bar\nu_{iL}(x)~\nu_{iR}(x) +\mathrm{h.c.})=-\sum^{3}_{i=1}m_{i}~\bar\nu_{i}(x)~\nu_{i}(x),
\end{equation}
where $\nu_{i}(x)=\nu_{iL}(x)+\nu_{iR}(x)$ is the  field of neutrino with the mass
\begin{equation}\label{YukawaNu6}
    m_{i}=y^{\nu}_{i}~v.
\end{equation}
It is easy to check that the Lagrangian of Standard Model with the Yukawa interaction (\ref{YukawaNu2})
is invariant under the global transformation
\begin{equation}\label{YukawaNu7}
  \nu_{iL}(x)\to e^{i\Lambda}\nu_{iL}(x),~~
 \nu_{iR}(x)\to e^{i\Lambda}\nu_{iR}(x),~~
  l_{L}(x)\to e^{i\Lambda}l_{L}(x),~~
  l_{R}(x)\to e^{i\Lambda}l_{R}(x),
\end{equation}
where $\Lambda$ is an arbitrary constant. Thus, in the Standard Model with massive neutrinos the total lepton number $L$ is conserved and {\em $\nu_{i}(x)$ is the Dirac field of neutrinos ($L=1$) and antineutrinos ($L=-1$).}

In spite Dirac neutrino masses can be introduced in the Standard Model,  this possibility looks extremely implausible.
Main reason is connected with the smallness of neutrino  masses and neutrino Yukawa couplings $y^{\nu}_{i}=\frac{m_{i}}{v}$.

Absolute values of neutrino masses at present   are unknown. However, from existing neutrino oscillation and cosmological data for the heaviest neutrino mass the following conservative bounds
can be found
\begin{equation}\label{YukawaNu11}
5\cdot 10^{-2}~\mathrm{eV}\simeq(\sqrt{\Delta m_{A}^{2}}) \leq m_{3}\leq (\frac{1}{3}\sum_{i}m_{i})\simeq 3\cdot 10^{-1}~\mathrm{eV},
\end{equation}
where $\Delta m_{A}^{2}\simeq 2.5\cdot 10^{-3}~\mathrm{eV}^{2}$
is the atmospheric neutrino mass-squared difference. From (\ref{YukawaNu11}) for the  Yukawa coupling $y^{\nu}_{3}$ we have
\begin{equation}\label{YukawaNu12}
    2\cdot 10^{-13}\leq y^{\nu}_{3}\leq 10^{-12}.
\end{equation}
Yukawa couplings of other particles of the third generation (t, b quarks and $\tau$-lepton) are equal, respectfully
\begin{equation}\label{YukawaNu10}
y_{t}\simeq 7\cdot 10^{-1},~~ y_{b}\simeq 2\cdot 10^{-2},~~y_{\tau}\simeq 7\cdot 10^{-3}.
\end{equation}
Thus,  Yukawa coupling of the heaviest neutrino {\em is more than ten orders of magnitude smaller} than Yukawa couplings of other particles of the third family.

Notice also that in the  SM Lagrangian, which do not include Yukawa interactions, enter left-handed and right-handed fields of all {\em charged particles}. Generation of their masses do not require additional degrees of freedom. The  $SU_{L}(2)\times U_{Y}(1)$ invariant SM can be built only with left-handed  neutrino fields.  The SM generation of neutrino masses {\em requires right-handed neutrino fields,  additional degrees of freedom.} Right-handed neutrino fields are out of line of economy and simplicity of the Standard Model.

 Thus, it is very unnatural and unlikely that neutrino masses and masses of leptons and quarks are of the same Standard Model origin. In the next section we will consider the most plausible (and popular)  beyond the Standard Model mechanism of the generation of neutrino masses.

\section{The Weinberg Effective Lagrangian Mechanism of the Neutrino Mass Generation}
 The Weinberg effective Lagrangian mechanism of the generation of small neutrino masses \cite{Weinberg:1979sa} is, apparently, the most popular beyond the SM mechanism. Before discussing this mechanism we will make the following remark.

In the framework of the approach, based on a neutrino mass term, neutrino masses and mixing were  introduced for the first time by Gribov and Pontecorvo in 1969 \cite{Gribov:1968kq}.\footnote{The results of this paper were generalized in \cite{Bilenky:1987ty}.}
At that time only $\nu_{e}$ and  $\nu_{\mu}$  were known and it was established that the lepton charged current had a $V-A$ form
\begin{equation}\label{CC}
    j^{CC}_{\alpha}=
    2(\bar\nu_{eL}\gamma_{\alpha}e_{L}+\bar\nu_{\mu L}\gamma_{\alpha}\mu_{L}).
\end{equation}
 In 1967 Pontecorvo's paper \cite{Ponte6} for the two lepton flavors all possible neutrino oscillations (between active neutrinos  and active and sterile neutrinos) were considered. Gribov and Pontecorvo  put the following question: is it possible to introduce neutrino masses and oscillations in the case if we assume that exist only left-handed neutrino fields $\nu_{eL}$ and  $\nu_{\mu L}$  and there are no right-handed sterile fields? The authors of the paper \cite{Gribov:1968kq} showed that it is possible to introduce neutrino masses in this case  {\em if the total lepton number $L$ is not conserved}.\footnote{At that time some authors claimed that if neutrinos are left-handed, their masses had to be equal to zero. This assertion is based, however, on the assumption that the total lepton number is conserved.}

Neutrinos have masses and are mixed if a neutrino mass term enters into the total Lagrangian. The  mass term is a sum of Lorenz-invariant products of left-handed and right-handed components. Authors of the paper \cite{Gribov:1968kq}
took into account  that the conjugated field
\begin{equation}\label{Cconj}
\nu_{lL}^{c}=C\bar\nu^{T}_{lL}
\end{equation}
   is right-handed. Here $C$ is the matrix of the charge conjugation which satisfies the relations $$C\gamma^{T}_{\alpha}C^{-1}=-\gamma_{\alpha},~~C^{T}=-C.$$
Thus, from the left-handed neutrino fields
$\nu_{lL}$ ($l=e,\mu,\tau$)  it was possible
to built the following neutrino mass term
\begin{equation}\label{Mj}
\mathcal{L}^{\mathrm{M}}= -\frac{1}{2}\,\sum_{l',l}
\bar\nu_{l'L}\,M^{\mathrm{M}}_{l'l}\nu^{c}_{lL}
+\mathrm{h.c.}
\end{equation}
where $M^{\mathrm{M}}$ is a complex, non diagonal $3\times 3$ matrix. From requirements of the Fermi-Dirac statistics it follows that $M^{\mathrm{M}}=(M^{\mathrm{M}})^{T}$.

Let us stress that
\begin{enumerate}
  \item The Majorana mass term $\mathcal{L}^{\mathrm{M}}$ is
{\em the only possible neutrino mass term} which can be built from the left-handed neutrino fields $\nu_{lL}$. This also means that the Majorana mass term is {\em the most economical general possibility} for neutrino masses and mixing (there are no right-handed neutrino fields in the Lagrangian, the number of neutrino degrees of freedom is minimal).

  \item The mass term $\mathcal{L}^{\mathrm{M}}$ does not conserve the total lepton number $L=L_{e}+L_{\mu}+L_{\tau}$.
 \end{enumerate}
The mass term $\mathcal{L}^{\mathrm{M}}$ is called {\em the Majorana mass term.}

The symmetrical matrix $M$ can be diagonalized by the following transformation
\begin{equation}\label{Mj1}
    M^{\mathrm{M}}=U~m~U^{T}.
\end{equation}
Here $U~U^{\dag}=1$ and $m_{ik}=m_{i}\delta_{ik},~~m_{i}>0$.
From (\ref{Mj}) and (\ref{Mj1}) we find
\begin{equation}\label{Mj2}
 \mathcal{L}^{\mathrm{M}}=
 -\frac{1}{2}\sum^{3}_{i=1}m_{i}~\bar\nu_{i}\nu_{i},
\end{equation}
where
\begin{equation}\label{Mj3}
\nu_{i}=\sum_{l}U_{il}^{\dag}\nu_{lL} +\sum_{l}(U_{il}^{\dag}\nu_{lL})^{c}.
\end{equation}
From (\ref{Mj2}) and (\ref{Mj3}) follows that
\begin{itemize}
  \item $\nu_{i}$ is the field of neutrino with the mass $m_{i}$.
  \item  The field $\nu_{i}$ satisfies the Majorana condition
 \begin{equation}\label{Mj4}
 \nu_{i}=\nu^{c}_{i}
 \end{equation}
 and  is  the Majorana field.
  \item
The flavor neutrino fields $\nu_{lL}$ are mixed fields
\begin{equation}\label{Mj5}
 \nu_{lL}=\sum^{3}_{i=1}U_{li}\nu_{iL}.
\end{equation}
\end{itemize}
The unitary $3\times3$ mixing matrix $U$ is called the PMNS matrix \cite{Pontecorvo:1957qd,Maki:1962mu}. Let us stress that
in the phenomenological Gribov-Pontecorvo approach {\em neutrino masses $m_{i}$ are parameters}.  There are no any explanation of the smallness of  these parameters.

 Weinberg \cite{Weinberg:1979sa} proposed a beyond the SM mechanism of neutrino mass generation which leads to the Majorana mass term and
 allows to explain smallness of neutrino masses. This mechanism is based on the effective Lagrangian approach.

If  SM particles interact with heavy, beyond SM particles with masses  much larger than  $v\simeq 246$ GeV,  then in the electroweak region fields of heavy particles can be "integrated out" and  this new interaction induce a nonrenormalizable interaction which is described by the effective Lagrangian. The effective Lagrangians are dimension five or more operators, invariant under   $SU_{L}(2)\times U_{Y}(1)$  transformations and  built from the Standard Model fields.

We are interested in the effective Lagrangian which generates a neutrino mass term.  We  assume that  only left-handed neutrino fields $\nu'_{lL}$,  components of the lepton doublets $\psi_{lL}$, enter in the SM.

Let us consider  $SU_{L}(2)\times U_{Y}(1)$  invariant
\begin{equation}\label{effL}
  (\tilde{\phi }^{\dag}~ \psi_{lL}) \quad (l=e,\mu,\tau),
\end{equation}
where $\tilde{\phi }=i\tau_{2}\phi^{*}$ is the conjugated Higgs doublet. After the spontaneous symmetry breaking for the proportional to $v$ term we have
\begin{equation}\label{effL1}
(\tilde{\phi }^{\dag}~ \psi_{lL})_{\mathrm{SSB}}= \frac{v}{\sqrt{2}}~\nu'_{lL}.
\end{equation}
From this expression it is obvious that   {\em the only possible effective Lagrangian} which generate neutrino mass term has the form \cite{Weinberg:1979sa}
\begin{equation}\label{effL2}
\mathcal{L}_{I}^{\mathrm{eff}}=-\frac{1}{\Lambda}~
\sum_{l',l}\overline{(\tilde{\phi }^{\dag} \psi_{l'L})}~ X'_{l'l}~(\tilde{\phi }^{\dag} \psi_{lL})^{c}+\mathrm{h.c.}=-\frac{1}{\Lambda}~\sum_{l',l}\bar \psi_{l'L}\tilde{\phi }~ X'_{l'l}~\tilde{\phi }^{\dag}~ \psi_{lL}^{c}
 +\mathrm{h.c.}
 \end{equation}
 where $X'$ is a symmetrical, non diagonal matrix.

 The operator in (\ref{effL2}) has a dimension $M^{5}$. Because a Lagrangian has dimension $M^{4}$,  the  parameter $\Lambda$, which has a dimesion $M$, is introduced in (\ref{effL2}). The parameter  $\Lambda$ characterizes a scale of a new physics (at  $\Lambda\to \infty $
effects of a new physics disappear). We could expect that
$\Lambda\gg v$.

Let us stress again that the operator in (\ref{effL2}) {\em is the only dimension five effective Lagrangian.} Other effective Lagrangians have dimension six and higher. In such effective Lagrangians   enter  coefficients
 $\frac{1}{\Lambda^{n}}$ with $n\geq 2$. This means that the {\em investigation of effects of neutrino masses and mixing}, ( neutrino oscillations, neutrinoless double $\beta$-decay etc) {\em is the most sensitive way to probe a beyond the SM new physics.}

Let us return back to the  effective Lagrangian (\ref{effL2}). After the spontaneous symmetry braking  we find the following neutrino mass term
\begin{equation}\label{effL3}
 \mathcal{L}^{\mathrm{M}}=-\frac{v^{2}}{2\Lambda}~\sum_{l',l}
 \bar\nu'_{l'L}~ X'_{l'l}~~ \nu'^{c}_{lL}
 +\mathrm{h.c.}
\end{equation}
 We can present $\mathcal{L}^{\mathrm{M}}$ in the standard form in which  flavor neutrino fields $\nu_{lL}$ enter. The SM leptonic charged current  has the following form
\begin{equation}\label{lepCC}
 j^{CC}_{\alpha}=2\sum_{l}\bar\psi_{lL}\frac{1}{2}
 (\tau_{1}+i\tau_{2})\gamma_{\alpha}\psi_{lL}
 =2\sum_{l}\bar\nu'_{lL}\gamma_{\alpha}l'_{L}.
\end{equation}
Taking into account (\ref{Yukawa7}) from (\ref{lepCC}) we find
\begin{equation}\label{lepCC1}
 j^{CC}_{\alpha}=2\sum_{l,l_{1}}\bar\nu'_{lL}\gamma_{\alpha}
 (V_{L})_{ll_{1}}l_{1L}=    2\sum_{l}\bar\nu_{lL}\gamma_{\alpha}l_{L}.
\end{equation}
Here $l$ is the field of the charged lepton with mass $m_{l}$ and the flavor (current) neutrino field $\nu_{lL}$ is given by
\begin{equation}\label{curnu}
    \nu_{lL}=\sum_{l_{1}}(V^{\dag}_{L})_{ll_{1}}\nu'_{l_{1}L}.
\end{equation}
From (\ref{effL3}) and (\ref{curnu})  we find the following Majorana mass term
\begin{equation}\label{effL4}
\mathcal{L}^{\mathrm{M}}= -\frac{1}{2}\,\sum_{l',l}
\bar\nu_{l'L}\,M^{\mathrm{M}}_{l'l}\nu^{c}_{lL}
+\mathrm{h.c.}
\end{equation}
in which the Majorana matrix
 $M^{\mathrm{M}}$    is given by the expression
\begin{equation}\label{effL5}
 M^{\mathrm{M}}=\frac{v^{2}}{\Lambda}~X,
\end{equation}
where $X=V^{\dag}_{L}~X'~(V^{\dag}_{L})^{T}=X^{T}$.

The symmetrical, dimensionless matrix $X$ can be presented in the form
\begin{equation}\label{effL6}
X=U~x~U^{T},
\end{equation}
where $U~U^{\dag}=1$  and $x_{ik}=x_{i}\delta_{ik},~~x_{i}>0$.

From (\ref{effL4}), (\ref{effL5}) and (\ref{effL6}) we find
\begin{equation}\label{effL7}
\mathcal{L}^{\mathrm{M}}=-\frac{1}{2}\sum^{3}_{i=1}m_{i}\bar \nu_{i}\nu_{i}.
\end{equation}
Here
\begin{equation}\label{effL8}
\nu_{i}(x)= \nu^{c}_{i}(x)=C~\bar\nu^{T}_{i}(x)
\end{equation}
is the field of the Majorana neutrino with the mass
\begin{equation}\label{effL9}
m_{i}=\frac{v}{\Lambda}(vx_{i})
\end{equation}
and
\begin{equation}\label{effL10}
\nu_{lL}=\sum^{3}_{i=1}U_{li}~\nu_{iL},
\end{equation}
where $U$ is the unitary PNMS mixing matrix.

Generated by the standard Higgs mechanism masses of quarks and leptons (and also $W^{\pm}$, $Z^{0}$ and Higgs bosons) are proportional to the Higgs vacuum expectation value $v$. This is obviously connected with the fact that $v$ is the only SM parameter  which has dimension of mass. The  Weinberg mechanism of the neutrino mass generation is characterized by two parameters with the dimension of mass: $v$ and $\Lambda$.  We have $m_{i}\to 0$ at $\Lambda \to \infty$. It is obvious also  that at $v \to 0$ neutrino masses disappear. Thus, for neutrino masses, generated by the effective Lagrangian mechanism, we naturally come to the expression (\ref{effL9}) from which it follows that generated by this mechanism neutrino masses are suppressed with respect to SM masses of quarks and leptons by the factor
\begin{equation}\label{effL1}
\frac{v}{\Lambda}=\frac{\mathrm{EW~scale}}{\mathrm{scale~of~a~new~ physics}},
\end{equation}
which is naturally much smaller than one.

From (\ref{effL9}) we can try to estimate the parameter $\Lambda$, which characterize the scale of a new physics. In accordance with latest neutrino oscillation and cosmological data let us assume hierarchy of neutrino masses. In this case we have
\begin{equation}\label{m3}
m_{3}\simeq(\sqrt{\Delta m_{A}^{2}})\simeq 5\cdot 10^{-2}~\mathrm{eV}.
\end{equation}
The parameters $x_{i}$ in (\ref{effL9}) are unknown. If we assume (by analogy with Yukawa couplings  of the particles of the third family) that $x_{3}\lesssim 1$  we obtain the following estimate
\begin{equation}\label{scale}
    \Lambda \lesssim 10^{15}~\mathrm{GeV}.
\end{equation}
The effective Lagrangian (\ref{effL2}) does not conserve the total lepton number $L$. Notice that the global invariance and conservation of $L$ (and the barion number $B$) are not  proper symmetries of the Quantum Field Theory (constant phases are not dynamical variable etc). In the Standard Model local gauge symmetry and renormalizable Yukawa interactions ensure conservation of $L$ and $B$ \cite{Weinberg:1980bf}. We could expect that a beyond the SM theory does not conserve $L$ and $B$ (see recent discussion in \cite{Witten:2017hdv}). This is an additional general argument in favor of Majorana nature of beyond the SM neutrino masses.

 In conclusion, let us stress again that we assumed (and this is our basic assumption) that in the Lagrangian of the minimal, renormalizable Standard Model there is no neutrino mass term. Then  in the framework of the non-renormalizable, beyond the Standard Model effective Lagrangian approach the residual $SU_{L}(2)\times U(1)_{R}$ symmetry  naturally ensure the smallness of neutrino masses(via the additional factor $\frac{v}{\Lambda}$ in the expression for the neutrino mass (\ref{effL9})).

\section{On the origin of the Weinberg Effective Lagrangian}
Let us consider the lepton number violating, $SU_{L}(2)\times U_{Y}(1)$ invariant interaction \cite{Weinberg:1979sa,Weinberg:1980bf}
\begin{equation}\label{seesaw1}
\mathcal{L}_{I}=-\sqrt{2}\sum_{l i}(\bar \psi_{l L}\tilde{\phi })y_{li}~N_{iR}+\mathrm{h.c.}
\end{equation}
Here
\begin{equation}\label{seesaw2}
N_{i}=N^{c}_{i}=C(\bar N_{i})^{T}, ~~~i=1,2,...n
\end{equation}
is the field of the Majorana heavy leptons, $SU_{L}(2)\times U_{Y}(1)$ singlet, $\psi_{l L}$ and $\tilde{\phi}$ are SM lepton and conjugated Higgs doublets and $y_{li}$ are Yukawa couplings.

If masses of the heavy leptons $M_{i}$ are much larger than $v$ the Lagrangian (\ref{seesaw1}) in the second order of the perturbation theory  generates the Weinberg effective Lagrangian. In fact, for the $S$-matrix we find
\begin{eqnarray}\label{seesaw3}
 S^{(2)}&=&\frac{(-i)^{2}}{2!}2\int T(\sum_{l',l, i,k}\bar \psi_{l' L}(x_{1})\tilde{\phi }(x_{1})y_{l'i}~N_{iR}(x_{1})\nonumber\\
 &\times& N^{T}_{kR}(x_{2})y_{lk}\tilde{\phi }^{T}(x_{2})\psi^{T}_{l L}(x_{2}))d^{4}x_{1}d^{4}x_{2}+...
\end{eqnarray}
In the electroweak region $Q^{2}\ll M_{i}^{2}$ for the heavy lepton's propagator we have
\begin{equation}\label{seesaw4}
 \langle 0|T(N_{iR}(x_{1})N^{T}_{kR}(x_{2}))|0\rangle\simeq i\frac{1}{M_{i}}
\delta(x_{1}-x_{2})\frac{1+\gamma_{5}}{2}C\delta_{ik}.
\end{equation}
From (\ref{seesaw3}) and (\ref{seesaw4})  we obtain the effective Lagrangian (\ref{effL2}) in which
\begin{equation}\label{seesaw5}
\frac{1}{\Lambda}X'_{l'l}=\sum_{i}y_{l'i}\frac{1}{M_{i}}y_{li}.
\end{equation}
From this relation it follows that a scale of a new physics $\Lambda$ is determined by masses of heavy Majorana leptons $N_{i}$.\footnote{Heavy Majorana leptons could explain the barion asymmetry of the Universe.  In fact, decays of $N_{i}$'s, produced  in the early Universe, into Higgs-lepton pairs could create  a lepton asymmetry, if the interaction (\ref{seesaw1}) violates $CP$. The lepton asymmetry via QCD sphaleron processes can generate the barion asymmetry. This mechanism of the generation of the barion asymmetry of the Universe is called leptogenesis (see reviews \cite{Buchmuller:2004tu,Davidson:2008bu,DiBari:2012fz}).}

Thus, the Weinberg effective Lagrangian  and (after spontaneous symmetry breaking) Majorana neutrino mass term can be generated by the exchange   of heavy virtual Majorana leptons, $SU_{L}(2)\times U_{Y}(1)$ singlets, between  lepton-Higgs pairs. This mechanism of the generation of Majorana neutrino masses and mixing is called type-I seesaw mechanism \cite{Minkowski:1977sc,GellMann:1980vs,Yanagida:1979as,Glashow:1979nm,
      Mohapatra:1980yp}

The interaction (\ref{seesaw1}) is not, however, the only possible interaction which generate the effective Lagrangian  (\ref{effL2}) in the tree-approximation. The Weinberg effective Lagrangian can be also generated in the tree-approximation by interaction of heavy triplet scalar boson field with pair of lepton doublet and pair of Higgs doublet fields. This scenario is called the type-II seesaw mechanism.

Finally, the Weinberg effective Lagrangian can be generated
in the tree approximation  by the exchange of heavy virtual Majorana triplet leptons  between  lepton-Higgs pairs. This mechanism is called the type-III seesaw mechanism.

\section{ General Remarks on the Neutrino Mass Generation }
The interaction (\ref{seesaw1}) and other seesaw interactions are  minimal (in a sense of  new degrees of freedom) beyond the SM, lepton number violating possibilities of generation of the effective Weinberg Lagrangian and small Majorana neutrino masses.
However,  the scale $\Lambda$ of a new physics  is naturally very large , unreachable in laboratory experiments ($\Lambda\lesssim 10^{15}$~GeV). This  circumstance inspired creations of many non minimal radiative neutrino mass models with   new physics at much
lower scales. Detailed discussion of such models and their classification can be found in the review \cite{Cai:2017jrq} where references to numerous original papers can be found.

A wide class of radiative models is based on the assumption
that  neutrinos in the Standard Model are massless, left-handed particles and that there are more new heavy fields and more interactions, than in the tree-level seesaw cases. The Weinberg effective Lagrangian is generated by interactions of SM fields and beyond the SM heavy fields via one or more loops. Smaller  $\Lambda$ then in the the seesaw cases are usually needed to explain values of neutrino masses. \footnote{One of the reason is that with every loop enters a suppression factor $\frac{1}{16\pi^{2}}\simeq 0.006$.}

In many radiative neutrino mass models dimension five Weinberg effective Lagrangian (\ref{effL2}) is generated. Some models lead to a dimension $5+2n$ effective Lagrangian
\begin{equation}\label{GenWein}
\mathcal{L}_{I}^{\mathrm{eff}}=-\frac{1}{\Lambda}~(\bar \psi_{L}\tilde{\phi })~ X'^{n}~(\tilde{\phi }^{\dag}~ \psi_{L}^{c})\frac{(\phi^{\dag}\phi)^{n}}{\Lambda^{2n}}
 +\mathrm{h.c.}
\end{equation}
Here $n=1,2,...$.  For the Majorana neutrino masses we have in this case
\begin{equation}\label{GenWein1}
    m_{i}= v~(\frac{v}{\Lambda})^{2n+1}x^{n}_{i}
\end{equation}
From this expression it follows that a scale of a new physics $\Lambda$, much smaller then  in the classical Weinberg case, is required in order to ensure smallness of neutrino masses.

Let us now try to extract some {\em general conclusions}  from a wide class of neutrino mass models, tree-level and radiative, which generate the Weinberg effective Lagrangian or its generalization  and are based on the following assumptions
\begin{itemize}
  \item The Standard Model neutrinos $\nu_{l}$ are massless left-handed particles (there are no right-handed neutrino fields in the SM).
  \item Small neutrino masses are generated by a $L$-violating, beyond a SM interactions and a scale of a new physics  is much larger than the electroweak scale $v$.
\end{itemize}
In the electroweak region  effects of beyond the SM particles are integrating out and the effective Weinberg Lagrangian (or its generalizations) and, correspondingly, the Majorana neutrino mass term
 \begin{equation}\label{1Mj}
\mathcal{L}^{\mathrm{M}}= -\frac{1}{2}\,\sum_{l',l=e,\mu,\tau}
\bar\nu_{l'L}\,M^{\mathrm{M}}_{l'l}\nu^{c}_{lL}
+\mathrm{h.c.}
\end{equation}
is generated.

Basic assumptions, listed above, ensure smallness of   Majorana neutrino masses with respect to the SM masses of leptons and quarks. However, values of neutrino masses  depend on unknown parameters and can not be predicted. {\em The common features of  all models, in which the Weinberg effective Lagrangian is generated}, are the following
\begin{enumerate}
  \item Neutrinos with definite masses $\nu_{i}$ are Majorana particles.
  \item The number of neutrinos with definite masses $\nu_{i}$ is equal to the number of the lepton flavors (three).
\end{enumerate}
The most sensitive experiments, which allow us to probe the Majorana nature of $\nu_{i}$, are experiments on the search for neutrinoless double $\beta$- decay  of some even-even nuclei. In the next sections we will briefly consider this process.

If the number of massive neutrinos if equal to three, it will be no {\em sterile neutrinos}, neutrinos which do not have the Standard Model weak interaction.  As it is well known,  indications in favor of sterile neutrinos  were obtained in different short baseline neutrino experiments: in the appearance $\bar\nu_{\mu}\to \bar\nu_{e}$ LSND  \cite{Aguilar:2001ty} and  $\nua{\mu}\to \nua{e}$ MiniBooNE  \cite{Aguilar-Arevalo:2018gpe} experiments, in the disappearance $\bar\nu_{e}\to \bar\nu_{e}$ reactor experiments \cite{Mention:2011rk} and in the disappearance $\nu_{e}\to \nu_{e}$  source Gallium experiments \cite{Giunti:2010zu}.

The existing data can be explained by neutrino oscillations if
we assume that fields of three flavor neutrinos $\nu_{e,\mu,\tau}$ and one sterile neutrino $\nu_{s}$ are mixtures  of the fields of four massive  neutrinos $\nu_{1,2,3,4}$  with $m_{1,2,3}\ll m_{4}$ (so called $3+1$ scheme). From analysis of the data follows that heavy mass $m_{4}$ is in the range $(10^{-1}\lesssim m_{4}\lesssim 10)$ eV.

Several new short baseline reactor, accelerator and source neutrino experiments are going on or in preparations at present. Analysis of the status of the light sterile neutrino was done in \cite{Dentler:2018sju,Boser:2019rta}. The latest update of the status of experiments on the search for sterile neutrinos can be found  in talks presented at the NEUTRINO 2020 conference (see
 http://nu2020.fnal.gov). From existing data it is not possible to make a definite conclusion on the existence of sterile neutrinos. I will limit myself by the following remarks:
\begin{enumerate}
  \item In new reactor experiments DANSS \cite{Shitov} (with reactor-detector distances 10.7, 11.7 and 12.7 m) and PROSPECT\cite{Andriamirado:2020erz} (with reactor-detector distances in the range 6.7-9.2 m) no indications in favor of short baseline neutrino oscillations were found. Best-fit point of previous reactor experiments is excluded in both experiments (at 5$\sigma$ in the DANSS experiment).
  \item Combined analysis of the data, obtained in the reactor Daya Bay and Bugey-3 experiments and accelerator MINOS+  experiment, allows to exclude at 90 \% CL LSND and MiniBooNE allowed regions for $\Delta m^{2}_{14}< 5~\mathrm{eV}^{2}$  \cite{Adamson:2020jvo}.
\item In the framework of $3+1$ oscillation scheme with one mass-squared difference for all oscillation channels   there exist a relation between oscillation amplitudes  which  allows to predict from LSND and old reactor data
 the oscillation amplitude of $\nu_{\mu}\to \nu_{\mu}$ transition \cite{Bilenky:1996rw,Okada:1996kw}. This prediction is in a strong tension with negative results of the search for short-baseline $\nua{\mu}\to \nua{\mu}$ oscillations in MINOS+ and IceCube experiments (see \cite{Gariazzo:2017fdh,Dentler:2018sju}).
Notice that this problem does not disappear  if  the data are analyzed in the framework of schemes with two or more sterile neutrinos.

\item   Sensitive searches for light sterile neutrinos will be performed in the nearest future.  In the SBN experiment at Fermilab \cite{Machado:2019oxb} three liquid argon TPC detectors at the distances 110 m, 470 m and 600 m from the target will be used. In the same experiment the search for oscillations in appearance $\nu_{\mu}\to \nu_{e}$  and disappearance $\nu_{\mu}\to \nu_{\mu}$ channels will be performed. It is planned that the full 99\% LSND allowed region will be covered at more than 5 $\sigma$. In the $JSNS^{2}$ experiment at J-PARC (Japan) \cite{Ajimura:2017fld} the direct test of the LSND result will be performed. Much better than in the LSND experiment beam and better detector (Gd-loaded liquid scintillator) will be used. It is expected that a background in the $JSNS^{2}$ experiment will be much lower than in the LSND experiment. The experiment with one detector would allow to exclude the larger part of the LSND-allowed region. In future experiment with two detectors all  LSND-allowed region will be covered. There are all reasons to believe that the sterile neutrino anomalies will be resolved in the nearest years.
\end{enumerate}

\section{On the Phenomenological Theory of the $0\nu\beta\beta$-decay}
Experiments on the search for the lepton number violating $0\nu\beta\beta$-decay of some even-even nuclei
\begin{equation}\label{bb1}
(A,Z)\to (A,Z+2)+e^{-}+e^{-}
\end{equation}
 ensure an unique probe of the Majorana nature of neutrinos with definite masses. These experiments have many advantages with respect to other possible $L$-violating experiments: large targets (in future experiments about 1 ton or more), law backgrounds, high energy resolutions etc. However, $0\nu\beta\beta$-experiments are extremely difficult and challenging. This is connected with the fact that the expected probabilities of the neutrinoless double $\beta$-decay of different nuclei (in the case if $\nu_{i}$ are Majorana particles) are extremely small. The main reasons are the following:
\begin{itemize}
  \item  The $0\nu\beta\beta$-decay is
   the second order of the perturbation theory in the Fermi constant $G_{F}\simeq 1.166\cdot10^{-5}~\frac{1}{\mathrm{GeV}^{2}}$ process.

  \item Because of the $V-A$ nature of the weak interaction  the matrix element of the $0\nu\beta\beta$-decay is proportional to the effective Majorana mass $m_{\beta\beta}=\sum_{k}U^{2}_{ek}m_{k}$. Smallness of the neutrino masses is a reason for the additional severe suppression of the probability of the decay.

\item There are two possibilities for neutrino mass spectra: Normal Ordering or Inverted Ordering (see later). Existing neutrino oscillation data favor Normal Ordering. In the case of the Normal Ordering the effective Majorana mass is much smaller than in the case of the Inverted Ordering.

\end{itemize}
The standard theory of the neutrinoless double $\beta$-decay is based on the following assumptions (see reviews \cite{Doi:1985,Vergados:2016hso,DellOro:2016tmg,
Bilenky:2014uka})
\begin{enumerate}
  \item The CC interaction is the SM interaction
 \begin{equation}\label{CCLag}
{\mathcal{L}}^{CC}_{I}(x) = -\frac{g}{2\sqrt{2}}\,j_{\alpha}^{CC}(x)
W^{\alpha}(x) + \mathrm{h.c.}.
\end{equation}
Here
\begin{equation}\label{CCcurrent}
j_{\alpha}^{CC}(x)=2\sum_{l=e,\mu,\tau} \bar \nu_{l L}(x)
\gamma_{\alpha} l_{L}(x)+j^{CC\mathrm{quark}}_{\alpha}(x).
\end{equation}
Here $j^{CC\mathrm{quark}}_{\alpha}(x)$ is the quark charged current, $W^{\alpha}(x)$ is the field of the charged $W^{\pm}$ vector bosons and $g$ is the constant of the electroweak interaction.
 \item The flavor neutrino field $\nu_{l L}(x)$ is given by the mixing relation
 \begin{equation}\label{mixture}
\nu_{lL}(x) = \sum^{3}_{k=1} U_{lk} \nu_{kL}(x),
\end{equation}
 where
\begin{equation}\label{mix}
\nu_{k}(x)=\nu^{c}_{k}(x)=C\bar\nu^{T}_{k}(x),\quad  (k=1,2,3)  \end{equation}
is the  field of {\em the Majorana neutrino} with mass $m_{k}$ and $U$ is the  unitary $3\times 3$ PMNS mixing matrix.\footnote{We assume that a beyond the SM physics contributes to the matrix element of the $0\nu\beta\beta$-decay via Majorana neutrino masses. If neutrinos with definite masses are Majorana particles their masses definitely contribute  to the matrix element of the $0\nu\beta\beta$-decay. For discussion of possible additional  contribution of  effective Lagrangians to the $0\nu\beta\beta$-decay see, for example, \cite{delAguila:2012nu}.}
\end{enumerate}
Existing weak interaction data are in perfect agreement with (\ref{CCLag}). All atmospheric, accelerator, solar and long baseline reactor  neutrino oscillation data are in agreement with the three neutrino mixing (\ref{mixture}).

The effective Hamiltonian of the $\beta$-decay is given by the expression
\begin{equation}\label{effham}
{\mathcal{H}}_{I}(x)= \frac{G_F}{\sqrt{2}} 2~\bar e_{L}(x)
\gamma_{\alpha}~ \nu_{eL}(x)~j^{\alpha}(x) + \mathrm{h.c.}
\end{equation}
where $j^{\alpha}(x)$ is $\Delta S=0$ hadronic charged current and  the field of electron neutrinos $\nu_{eL}(x)$ is given by (\ref{mixture}).

In the second order of the perturbation theory in $G_{F}$ the matrix element of the $0\nu\beta\beta$-decay is given by the following expression
\begin{eqnarray}\label{Smatelem}
&&\langle f|S^{(2)}|i\rangle=4\frac{(-i)^{2}}{2~!}~\left (\frac{G_F}{\sqrt{2}}\right )^{2}N_{p_1}N_{p_2}
\int\bar u_{L}(p_1)e^{ip_{1}x_{1}}\gamma_{\alpha}~\langle 0|T(\nu_{eL}(x_{1})~\nu^{T}_{eL}(x_{2})|0\rangle
\nonumber\\
&&\times \gamma^{T}_{\beta}~\bar
u^{T}_{L}(p_2)e^{ip_{2}x_{2}}\langle
N_{f}|T(J^{\alpha}(x_{1})J^{\beta}(x_{2}))|N_{i}\rangle
~d^{4}x_{1}d^{4}x_{2}-(p_{1}\leftrightarrows p_{2}).
\end{eqnarray}
Here $p_{1}$ and $p_{2}$ are electron momenta, $J^{\alpha}(x)$
is the weak charged current in the Heisenberg representation, $N_{i}$ ($N_{f}$) are  states of the initial (final) nuclei
with 4-momenta $P_{i}=(E_{i}, \vec{p}_{i})$ ($P_{f}=(E_{f}, \vec{p}_{f})$) and $N_{p}=\frac{1}{(2\pi)^{3/2}\sqrt{2p^{0}}}$ is the standard normalization factor.

From the Majorana condition (\ref{mix}) follows that
\begin{equation}\label{mix1}
\nu^{T}_{k}(x)=\bar\nu_{k}(x)C^{T}=-\bar\nu_{k}(x)C.
\end{equation}
Using this relation, for the neutrino propagator we have
\begin{eqnarray}\label{nuprop}
&&\langle 0|T(\nu_{eL}(x_{1})~\nu^{T}_{eL}(x_{2})|0\rangle=
-\sum_{k}U^{2}_{ek}\frac{1-\gamma_{5}}{2}\langle 0|T(\nu_{k}(x_{1})~\bar\nu_{k}(x_{2})|0\rangle\frac{1-\gamma_{5}}{2}C
\nonumber\\&&=-\frac{i}{(2\pi)^{4}}\sum_{k} U^{2}_{ek}
\int\frac{m_{k}~e^{-iq(x_{1}-x_{2})}}{q^{2}-m^{2}_{k}} d^{4}q~\frac{1-\gamma_{5}}{2}C.
\end{eqnarray}
Further, taking into account that
\begin{equation}\label{nuprop1}
\bar u_{L}(p_1)\gamma_{\alpha}\frac{1-\gamma_{5}}{2}\gamma_{\beta}C~\bar
u^{T}_{L}(p_2)=- \bar u_{L}(p_2)\gamma_{\beta}\frac{1-\gamma_{5}}{2}\gamma_{\alpha}C~\bar
u^{T}_{L}(p_1)
\end{equation}
and
\begin{equation}\label{nuprop2}
T(J^{\alpha}(x_{1})J^{\beta}(x_{2}))
=T(J^{\beta}(x_{2})J^{\alpha}(x_{1}))
\end{equation}
from (\ref{Smatelem}) and (\ref{nuprop}) for the matrix element of
$0\nu\beta\beta$-decay we obtain the following expression
\begin{eqnarray}\label{Smatelem1}
\langle f|S^{(2)}|i
\rangle&=&-4 \left (\frac{G_F}{\sqrt{2}}\right )^{2}N_{p_1}N_{p_2}
 \int\bar
u_{L}(p_1)e^{ip_{1}x_{1}}\gamma_{\alpha}\left(\frac{i}{(2\pi)^{4}}
U^{2}_{ek}\sum_{k}\int \frac{m_{k}~e^{-iq~(x_{1}-x_{2})}}
{ q^{2}-m^{2}_{k}}d^{4}q\right)\nonumber\\
&&\times \gamma_{\beta}\frac{1+\gamma_{5}}{2}C~\bar
u^{T}_{L}(p_2)e^{ip_{2}x_{2}} \langle
N_{f}|T(J^{\alpha}(x_{1})J^{\beta}(x_{2}))|N_{i}
\rangle~d^{4}x_{1}d^{4}x_{2}.
\end{eqnarray}
In this expression we can perform integration over $x^{0}_{1}$, $x^{0}_{2}$ and $q^{0}$. The matrix element of the $0\nu\beta\beta$-decay takes the form
\begin{eqnarray}\label{Smatelem2}
&&\langle f|S^{(2)}|i
\rangle=2i~\left(\frac{G_F}{\sqrt{2}}\right )^{2}N_{p_1}N_{p_2}
 \bar
u(p_1)\gamma_{\alpha}\gamma_{\beta}(1+\gamma_{5})C \bar
u^{T}(p_2)\int
d^{3}x_{1}d^{3}x_{2}e^{-i\vec{p}_{1}\vec{x}_{1}-i\vec{p}_{2}\vec{x}_{2}}
 \nonumber\\
&&\times\sum_{k}U^{2}_{ek}m_{k}\frac{1}{(2\pi)^{3}}
\int\frac{e^{i\vec{q}~(\vec{x}_{1}-\vec{x}_{2})}} {
q_{k}^{0}}d^{3}q~[\sum_{n} \frac{\langle
N_{f}|J^{\alpha}(\vec{x}_{1})|N_{n}\rangle\langle N_{n}|
J^{\beta}(\vec{x}_{2}))|N_{i}\rangle }{E_{n}+p^{0}_{2}+q^{0}_{k}-E_{i}-i\epsilon}
\nonumber\\ &&+\sum_{n}\frac{ \langle
N_{f}|J^{\beta}(\vec{x}_{2})|N_{n}\rangle\langle N_{n}|
J^{\alpha}(\vec{x}_{1}))|N_{i}\rangle}{E_{n}+p^{0}_{1}+q^{0}_{k}-E_{i}-i\epsilon}
]~2\pi\delta(E_{f}+p^{0}_{1}+p^{0}_{2}-E_{i}).
\end{eqnarray}
Here $q^{0}_{k}=\sqrt{\vec{q}^{2}+m_{k}^{2}}$  and $|N_{n}\rangle$ is the vector of the state of the intermediate nucleus with 4-momentum $P_{n}=(E_{n},\vec{p}_{n}$). In (\ref{Smatelem2}) the sum over the total system of the states $|N_{n}\rangle$ is assumed. Notice that  we used the relation
\begin{equation}\label{rel}
 J^{\alpha}(x)=e^{iHx^{0}}J^{\alpha} (\vec{x})e^{-iHx^{0}},\quad  J^{\alpha} (\vec{x})= J^{\alpha} (0,\vec{x}).
\end{equation}
The equation (\ref{Smatelem2}) is an exact expression for the matrix element of the $0\nu\beta\beta$-decay in the second order of the perturbation theory. The following approximations are standard ones
\begin{enumerate}
  \item  Small neutrino masses $m_{k}$  in the expression for the neutrino energy can be neglected ($q^{0}_{k}\simeq |\vec{q}|=q$). In fact, an average neutrino momentum is given by  $\bar{q}\simeq \frac{1}{r}$, where $r\simeq 10^{-13}$ cm is the average distance between nucleons in a nucleus. We have
 $\bar{q}\simeq 100~\mathrm{MeV}\gg m_{k}$.

\item Long-wave approximation
     $e^{-i\vec{p}_{i}\vec{x}_{i}}\simeq 1$~($i=1,2$). In fact, we have $|\vec{p}_{i}\cdot
\vec{x}_{i}|\leq p_{i}R$, where $R\simeq 1.2\cdot10^{-13}~A^{1/3}~\mathrm{cm}$ is the radius of a nucleus. Taking into account that $p_{i}\lesssim 1 $ MeV we find $|\vec{p}_{i}\cdot\vec{x}_{i}|\ll 1$.

\item Closure approximation. The energy of neutrino in an intermediate state $q\simeq 100$ MeV is much larger than the excitation energy $(E_{n}-E_{i})$. This means that energies of intermediate states $E_{n}$ in the denominators of Eq.(\ref{Smatelem2}) can be replaced by an average energy $\bar E$. In this approximation we can sum over intermediate states $|N_{n}\rangle$ in (\ref{Smatelem2}).
\end{enumerate}
In the laboratory frame for the energy denominators in (\ref{Smatelem2}) we have
\begin{equation}\label{rel1}
E_{n}+p^{0}_{1,2}+q^{0}_{k}-E_{i}\simeq \bar E+(\frac{p^{0}_{1}+p^{0}_{2}}{2})\pm (\frac{p^{0}_{1}-p^{0}_{2}}{2})+q-M_{i}\simeq \bar E+q-\frac{M_{i}+M_{f}}{2},
\end{equation}
where $M_{i}(M_{f})$ is the mass if initial (final) nucleus. In (\ref{rel1}) we neglect nuclear recoil and take into account that $(\frac{p^{0}_{1}-p^{0}_{2}}{2})\ll q$. For the matrix element of the $0\nu\beta\beta$-decay we find the following expression
\begin{eqnarray}\label{Smatelem3}
&&\langle f|S^{(2)}|i
\rangle\simeq2i~\left(\frac{G_F}{\sqrt{2}}\right )^{2}N_{p_1}N_{p_2}
 \bar
u(p_1)\gamma_{\alpha}\gamma_{\beta}(1+\gamma_{5})C \bar
u^{T}(p_2) m_{\beta\beta}
 \nonumber\\
&&\times \int
d^{3}x_{1}d^{3}x_{2}\frac{1}{(2\pi)^{3}}
\int\frac{e^{i\vec{q}~(\vec{x}_{1}-\vec{x}_{2})}} {
q(\bar E+q-\frac{M_{i}+M_{f}}{2})}d^{3}q~[ \langle
N_{f}|(J^{\alpha}(\vec{x_{1}})
J^{\beta}(\vec{x_{2}}) \nonumber\\
&&+J^{\beta}(\vec{x_{2}})
J^{\alpha}(\vec{x_{1}}))|N_{i}\rangle
]~2\pi\delta(E_{f}+p^{0}_{1}+p^{0}_{2}-E_{i}),
\end{eqnarray}
where
\begin{equation}\label{eMjmass}
 m_{\beta\beta}=\sum^{3}_{i=1}U^{2}_{ek}~m_{k}
\end{equation}
is {\em the effective Majorana mass}.

Thus, due to the smallness of neutrino masses the matrix element of the $0\nu\beta\beta$-decay is a product the effective Majorana mass $m_{\beta\beta}$ (which depends on absolute values of neutrino masses with unknown lightest neutrino mass and on (known) $\theta_{13}$ and  $\theta_{12}$ and on (unknown) Majorana phases) and nuclear part  (which include the propagator of virtual neutrino). As a result of this factorization, the total decay rate of the $0\nu\beta\beta$-decay has the following general form
\begin{equation}\label{totrate}
\Gamma^{0\nu}=\frac{\ln 2}{T^{0\nu}_{1/2}}=|m_{\beta\beta}|^{2}~
|M^{0\nu}|^{2}~G^{0\nu}(Q,Z).
\end{equation}
Here $M^{0\nu}$ is the nuclear matrix element (NME) and $G^{0\nu}(Q,Z)$ is the known phase-space factor which includes effects of the Column interaction of electrons and daughter nucleus.

Calculation of the nuclear  matrix elements of the $0\nu\beta\beta$-decay is a complicated many-body nuclear problem. Different approximate  methods are used in such calculations.
At the moment the results of different calculations of NME differ by 2-3 times. Discussion of these calculations
is out of the scope of this review (see reviews\cite{Vergados:2016hso,Engel:2016xgb,Menendez:2020}).

\section{Effective Majorana Mass}
In the case of the three neutrino mixing there are two independent mass-squared differences. From analysis of experimental data it follows that one mass-squared difference is about 30 times smaller than the other one. The small (large) mass-squared difference is usually called solar (atmospheric) and is denoted by $\Delta m_{S}^{2}$ ($\Delta m_{A}^{2}$).

Neutrino masses are labeled in such a way  that  the solar mass-squared difference is given by\footnote{We will use the following definition: $\Delta m^{2}_{ki}= m^{2}_{i}- m^{2}_{k}$.}
\begin{equation}\label{sol}
\Delta m^{2}_{S}\equiv\Delta m^{2}_{12}.
\end{equation}
From the MSW effect \cite{Wolfenstein:1977ue,Mikheev:1986wj}, observed in solar neutrino experiments, follows that\footnote{In fact, the MSW resonance condition has a form $\Delta m^{2}_{12}\cos2\theta_{12}=2\sqrt{2}G_{F}n_{e}E>0$ ($n_{e}$ is the electron number density). From this condition follows that $\Delta m^{2}_{12}>0$  under the standard assumption that the principal values of the mixing angle $\theta_{12}$ are in the range $0\leq \theta_{12}\leq \frac{\pi}{2}$.}
\begin{equation}\label{sol}
\Delta m^{2}_{12}>0.
\end{equation}
For the mass of the third neutrino $m_{3}$ there are two possibilities:
\begin{enumerate}
  \item  Normal ordering (NO)
 \begin{equation}\label{NO}
    m_{3}>m_{2}>m_{1}
 \end{equation}
\item Inverted ordering (IO)
\begin{equation}\label{IO}
    m_{2}>m_{1}>m_{3}.
 \end{equation}
\end{enumerate}
Atmospheric mass-squared difference $\Delta m^{2}_{A} $ can be determined as follows\footnote{Notice that other definitions of $\Delta m^{2}_{A} $  are also used in the literature (see \cite{Bilenky:2015bvt}).}
\begin{equation}\label{deltaA}
\Delta m^{2}_{A}=\Delta m^{2}_{23} ~(NO),\quad
\Delta m^{2}_{A}=|\Delta m^{2}_{13}| ~(IO).
\end{equation}
In NO case for the neutrino masses $m_{2}$ and $m_{3}$ we obviously have
\begin{equation}\label{norspec1}
m_{2}=\sqrt{m^{2}_{1}+\Delta m^{2}_{S}},~~
m_{3}=\sqrt{m^{2}_{1}+\Delta m^{2}_{S}+\Delta m^{2}_{A}}.
\end{equation}
In the case of IO we find
\begin{equation}\label{invspec1}
m_{1}=\sqrt{m^{2}_{3}+\Delta m^{2}_{A}},\quad
m_{2}=\sqrt{m^{2}_{3}+\Delta m^{2}_{A}+\Delta m^{2}_{S}}
\end{equation}
If neutrinos with definite masses   $\nu_{i}$ are Dirac particles the PNMS mixing matrix  is characterized by three mixing angles
$\theta_{12}, \theta_{23}, \theta_{13}$
and one phase $\delta$. In the standard parametrization the mixing matrix  has the form
\begin{eqnarray}
U^{\mathrm{D}}=\left(\begin{array}{ccc}c_{13}c_{12}&c_{13}s_{12}
&s_{13}e^{-i\delta}\\
-c_{23}s_{12}-s_{23}c_{12}s_{13}e^{i\delta}&
c_{23}c_{12}-s_{23}s_{12}s_{13}e^{i\delta}&c_{13}s_{23}\\
s_{23}s_{12}-c_{23}c_{12}s_{13}e^{i\delta}&
-s_{23}c_{12}-c_{23}s_{12}s_{13}e^{i\delta}&c_{13}c_{23}
\end{array}\right).
\label{unitmixU1}
\end{eqnarray}
Here  $c_{ik}=\cos\theta_{ik},~s_{ik}=\sin\theta_{ik}$. If in the lepton sector $CP$ is conserved $U^{\mathrm{D}*}=U^{\mathrm{D}}$ and $\delta=0$.

If $\nu_{i}$ are Majorana particles, the mixing matrix is characterized by three angles and three phases and has the form
\begin{equation}\label{3Mjmix}
    U^{M}=U^{\mathrm{D}}~S^{M}(\alpha),
\end{equation}
where
\begin{eqnarray}\label{3Mjcol}
 S^{M}(\alpha)=\left(
\begin{array}{ccc}
e^{i\alpha_{1}}&0&0\\
0&e^{i\alpha_{2}}&0\\
0&0&1\\
\end{array}
\right).
\end{eqnarray}
Values of $\Delta m^{2}_{S}$ and $\Delta m^{2}_{A}$, three neutrino mixing angles and $CP$ phase,  obtained from the  global fit of the neutrino oscillation data,
are presented in the Table I.
\begin{center}
 Table I
\end{center}
\begin{center}
 Values of neutrino oscillation parameters, obtained from the
global fit of the data \cite{Esteban:2018azc}
\end{center}
\begin{center}
\begin{tabular}{|c|c|c|}
  \hline  Parameter &  Normal Ordering& Inverted Ordering\\
\hline   $\sin^{2}\theta_{12}$& $0.310^{+0.013}_{-0.012}$& $0.310^{+0.013}_{-0.012}$
\\
\hline    $\sin^{2}\theta_{23}$& $0.582^{+0.015}_{-0.019}$& $ 0.582^{+0.015}_{-0.018}$
\\
\hline   $\sin^{2}\theta_{13}$ & $ 0.02240^{+0.00065}_{-0.00066}$&  $0.02263^{+0.00065}_{-0.00066}$
\\
\hline   $\delta $~(in $^{\circ}$) & $(217^{+40}_{-28})$& $ (280^{+25}_{-28})$
\\
\hline $\Delta m^{2}_{S}$& $(7.39^{+0.21}_{-0.20})\cdot 10^{-5}~\mathrm{eV}^{2}$&$(7.39^{+0.21}_{-0.20})\cdot 10^{-5}~\mathrm{eV}^{2}$\\
\hline $\Delta m^{2}_{A}$& $(2.525^{+0.033}_{-0.031})\cdot 10^{-3}~\mathrm{eV}^{2}$&$(2.512^{+0.034}_{-0.031})\cdot 10^{-3}~\mathrm{eV}^{2}$\\
\hline
\end{tabular}
\end{center}
Notice that from data of all the latest neutrino oscillation experiments follows that the preferable neutrino mass spectrum is the NO spectrum (see talks at  Neutrino 2020, http://nu2020.fnal.gov).

The effective Majorana mass strongly
depends on the character of the neutrino mass spectrum. For illustration we will consider the following  viable  neutrino mass spectra
\begin{center}
  {\bf Hierarchy of the neutrino masses}
\end{center}
\begin{equation}\label{hierar}
m_{1} \ll m_{2} \ll m_{3}.
\end{equation}
In this case we have
\begin{equation}\label{hierar1}
 m_{2}\simeq \sqrt{ \Delta
m^{2}_{S}},\quad m_{3}\simeq  \sqrt{ \Delta m^{2}_{A}}\quad
m_{1} \ll \sqrt{\Delta m^{2}_{S}}.
\end{equation}
Thus, in the case of the neutrino mass hierarchy neutrino masses $m_{2}$ and $m_{3}$ are determined by the solar and atmospheric mass-squared differences and lightest mass $m_{1}$ is very small. Neglecting its contribution  and using the standard parametrization of the PMNS mixing matrix  for the effective Majorana mass we have
\begin{equation}\label{hierar2}
|m_{\beta\beta}|\simeq\left |\,\cos^{2}\theta_{13}
 \sin^{2} \theta_{12}\, \sqrt{\Delta m^{2}_{S}} + e^{2i\,\alpha}
 \sin^{2} \theta_{13}\, \sqrt{\Delta m^{2}_{A}}\,\right |~,
\end{equation}
where $\alpha$ is a phase difference.

The first term in (\ref{hierar2}) is small because of the
smallness of $\Delta m^{2}_{S}$. The contribution to $|m_{\beta\beta}|$ of the ``large'' atmospheric mass-squared difference
$\Delta m^{2}_{A}$  is suppressed by
the smallness of $\sin^{2} \theta_{13} $. As a result,
 absolute values of the first and second terms in (\ref{hierar2}) are of the same order of magnitude. From (\ref{hierar2}) and Table I we find the  $3\sigma$ upper bound
\begin{equation}\label{globfit2}
|m_{\beta\beta}|\leq 4.29 \cdot 10^{-3}\rm{eV},
\end{equation}
which is much smaller than upper bounds of $|m_{\beta\beta}|$ reached in modern experiments on the search for the $0\nu\beta\beta$-decay (see later).

\begin{center}
  {\bf Inverted hierarchy of neutrino masses}
\end{center}
\begin{equation}\label{invhierar}
m_{3} \ll m_{1} < m_{2}.
\end{equation}
In this case from (\ref{invspec1})  we have
\begin{equation}\label{invhierar1}
m_{1}\simeq \sqrt{\Delta m^{2}_{A}},~~m_{2}\simeq\sqrt{ \Delta m^{2}_{A}}~(1+\frac{\Delta m^{2}_{S}}{2\, \Delta m^{2}_{A}})\simeq\sqrt{ \Delta m^{2}_{A}},~~m_{3}\ll\sqrt{
\Delta m^{2}_{A}}.
\end{equation}
The lightest mass $m_{3}$ in the expression for the effective Majorana mass is multiplied by the small parameter
$\sin^{2}\theta_{13}$. Neglecting the contribution of this term, we find
\begin{equation}\label{invhierar2}
|m_{\beta\beta}|\simeq \sqrt{ \Delta m^{2}_{A}}\cos^{2}\theta_{13}~ (1-\sin^{2}
2\,\theta_{12}\,\sin^{2}\alpha)^{\frac{1}{2}},
\end{equation}
where $\alpha=\alpha_{2}-\alpha_{1}$ is  the Majorana phase difference, the only unknown parameter in the expression (\ref{invhierar2}).

From (\ref{invhierar2}) we find the following upper and lower bounds for the effective Majorana mass
\begin{equation}\label{invhierar3}
\cos^{2}\theta_{13}~\cos  2\,\theta_{12} \,\sqrt{ \Delta m^{2}_{A} } \leq
|m_{\beta\beta}| \leq\cos^{2}\theta_{13}~\sqrt{ \Delta m^{2}_{A}},
\end{equation}
which are realized if there is $CP$ invariance in the lepton sector.

From  (\ref{invhierar3}) and Table I we find the following $3~\sigma$ range for the effective Majorana mass in the case of the inverted hierarchy of neutrino masses
\begin{equation}\label{invhierar4}
1.44\cdot 10^{-2}~\mathrm{eV}\leq  |m_{\beta\beta}|\leq 5.01\cdot 10^{-2}~\mathrm{eV}
\end{equation}
As we will see in the next section, future experiments  will be sensitive to the inverted hierarchy region.

\section{Experiments on the Search for $0\nu\beta\beta$-decay}
Up to now neutrinoless double $\beta$-decay  was not observed. The results of some recent experiments on the search for
$0\nu\beta\beta$-decay  are presented in the Table II (see \cite{Giuliani:2019uno,Agostini:2019hzm} ).  In the third column of the Table II the 90\% CL lower bounds for the half-live of the decay of different elements are given. In the fourth column  upper bound ranges for the effective Majorana mass $|m_{\beta\beta}|$ are presented. It was assumed that $g_{A}=1.27$ ($g_{A}$ is the axial constant)  and  nuclear matrix elements, calculated in different models, were taken from \cite{Engel:2016xgb}.

 \begin{center}
Table II
\end{center}
\begin{center}
Lower limits of half-lives and upper limits for the effective Majorana mass, obtained in recent experiments on the search for the $0\nu\beta\beta$-decay
\end{center}
\begin{center}
\begin{tabular}{|c|c|c|c}
  \hline experiment& nucleus& $T_{1/2}$($10^{25}$yr)&  $|m_{\beta\beta}|$ (eV) \\
\hline GERDA \cite{Agostini:2019hzm}&$^{76}$ Ge & 9& 0.10-0.23 \\
 \hline KamLAND-Zen \cite{KamLAND-Zen:2016pfg} & $^{136}$ Xe& 10.7& 0.08-0.24 \\
  \hline EXO-200 \cite{Albert:2017owj} & $^{136}$ Xe&1.8& 0.09-0.29   \\
\hline CUORE \cite{Adams:2019jhp}&$^{130}$ Te&3.2& 0.08-0.35\\
\hline
\hline CUPID-0 \cite{Azzolini:2018dyb}&$^{82}$ Se&0.24& 0.39-0.81\\
\hline
\end{tabular}
\end{center}
Many new experiments on the search for the $0\nu\beta\beta$-decay are in preparation (see \cite{Giuliani:2019uno,Detwiler}).
In these experiments the inverted hierarchy region and, possibly, part of the normal hierarchy region will be probed. We will mention only a few of them. In the KamLAND-Zen experiment ($^{136}$ Xe ) after 5 years of running the sensitivity  $T_{1/2}>2\cdot 10^{27}$yr will be reached. In the SNO+ ($^{130}$ Te), LEGEND ($^{76}$ Ge), n-EXO ($^{136}$ Xe ), CUPID ($^{100}$ Mo ) and NEXT-HD ($^{136}$ Xe ) sensitivities $T_{1/2}>1\cdot 10^{27}$yr, $T_{1/2}>2\cdot 10^{28}$yr, $T_{1/2}>5.7\cdot 10^{27}$yr, $T_{1/2}>1.1
\cdot 10^{27}$yr and $T_{1/2}>3\cdot 10^{27}$yr will be reached after 10 years of running.

Future experiments on the search for the $0\nu\beta\beta$-decay are planned to solve the most fundamental problem of the modern neutrino physics: {\em Are neutrinos with definite masses  Majorana or Dirac particles?}

\section{Conclusion}
Neutrinos   are the only fundamental fermions which can be Dirac or Majorana particles. If neutrino masses and mixing are generated by the Standard Higgs Mechanism in this case into the SM Lagrangian enter right-handed neutrino fields $\nu_{lR}$, the total lepton number $L$ is conserved and neutrino with definite masses  are Dirac particles. We know, however, that neutrino masses are many orders of magnitude  smaller then masses  of other fundamental fermions.  It is very unlikely that neutrinos masses are generated by the same Higgs mechanism as masses of leptons and quarks. {\em It is very plausible  that Standard Model neutrinos are massless particles  and neutrino masses are generated by a beyond the SM mechanism.}

The most economical and general possibility to introduce neutrino masses and mixing  is based on the assumption that there exist only left-handed flavor neutrino fields $\nu_{lL}$ ($l=e,\mu,\tau$). Neutrino masses can be introduced in this case only if the total lepton number $L$ is not conserved and neutrinos with definite masses $\nu_{i}$ ($i=1,2,3$) are  Majorana particles (Majorana mass term).

The Weinberg effective Lagrangian (and its generalizations), which are based on numerous beyond the SM models, lead to
the Majorana neutrino mass term, the three-neutrino mixing $$\nu_{lL}=\sum^{3}_{i=1}U_{li}\nu_{iL},\quad \nu_{i}=\nu^{c}_{i}$$
and neutrino masses $m_{i}$, which are naturally suppressed with respect to SM masses of leptons and quarks, by a  ratio of the electroweak scale $v$ and a scale of a new, beyond the SM physics $\Lambda$.

Non conservation of the lepton number, which is the basics  of such explanation of the smallness of neutrino masses, is an attractive and natural feature of a beyond the SM physics (see \cite{Witten:2017hdv}).

In spite of very strong arguments in favor of the Majorana nature of neutrinos, a possibility of the Dirac neutrinos is not excluded {\em experimentally}. Notice that  even a beyond the SM seesaw mechanism for the Dirac neutrinos, which could explain the smallness of neutrino masses, was proposed (see \cite{Cai:2017jrq}). {\em Observation of the neutrinoless double $\beta$-decay} and a proof that neutrinos with definite masses are Majorana particle would be unique and crucial test of our basic understanding of the phenomenon of small neutrino masses.

\end{document}